\documentclass[preprint,journal]{vgtc} 

\ifpdf
  \pdfoutput=1\relax                   
  \usepackage{graphicx}                
  \DeclareGraphicsExtensions{.pdf,.png,.jpg,.jpeg} 
\else
  \usepackage{graphicx}                
  \DeclareGraphicsExtensions{.eps}     
\fi%

\graphicspath{{figures/}{pictures/}{images/}{./}} 

\usepackage{microtype}                 
\PassOptionsToPackage{warn}{textcomp}  
\usepackage{textcomp}                  
\usepackage{mathptmx}                  
\usepackage{times}                     
\usepackage{cite}                      
\usepackage{tabu}                      
\usepackage{booktabs}                  
\usepackage{xcolor}

\newcommand{\forreview}[1]{{\color{black} #1}}

\ieeedoi{10.1109/TVCG.2019.2934285}
\onlineid{1172}

\vgtccategory{Research}
\vgtcpapertype{Design Study}

\title{Visualizing a Moving Target: A Design Study on Task Parallel Programs
in the Presence of Evolving Data and Concerns}

\author{Katy Williams, Alex Bigelow, and Katherine E. Isaacs}
\authorfooter{
\item
 Katy Williams, Alex Bigelow, and Katherine Isaacs are with the University of
 Arizona. E-mails: \{kawilliams,alexrbigelow,kisaacs\}@email.arizona.edu.
}

\shortauthortitle{Williams \MakeLowercase{\textit{et al.}}: Visualizing a
Moving Target}

\abstract{\forreview{Common pitfalls in visualization projects include lack of data availability and
  the domain users' needs and focus changing too rapidly for the design process
  to complete. While it is often prudent to avoid such projects, we argue it
  can be beneficial to engage them in some cases as the visualization process
  can help refine data collection, solving a ``chicken and egg'' problem of
  having the data and tools to analyze it. We found this to be the case in the
  domain of task parallel computing where such data and tooling is an open
  area of research. Despite these hurdles, we conducted a design study.
  Through a tightly-coupled iterative design process, we built Atria, a multi-view execution graph
  visualization to support performance analysis. Atria simplifies the initial
  representation of the execution graph by aggregating nodes as related to
  their line of code. We deployed Atria on multiple platforms, some requiring
  design alteration. We
  describe how we adapted the design study methodology to the
  ``moving target'' of both the data and the domain experts' concerns and how
  this movement kept both the visualization and programming project healthy.
  We reflect on our process and discuss what factors allow the project to be
  successful in the presence of changing data and user needs.}
} 


\CCScatlist{ 
 \CCScat{K.6.1}{Management of Computing and Information Systems}%
{Project and People Management}{Life Cycle};
 \CCScat{K.7.m}{The Computing Profession}{Miscellaneous}{Ethics}
}

\teaser{
  \centering
  \includegraphics[width=\linewidth]{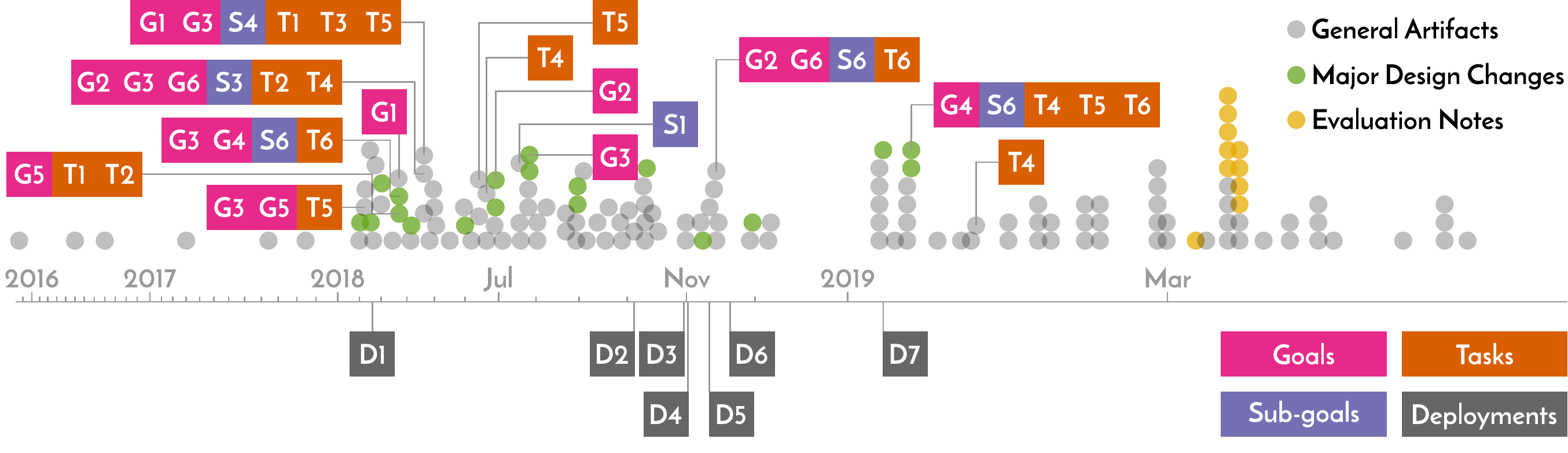}
  \caption{\forreview{Design Study timeline (log scale). The top contains a mark for each collected
  artifact. 
  Connections to identified goals, sub-goals, and tasks are marked when direct evidence for them has been identified. 
  Artifacts from meetings presenting major design changes and notes
  from the evaluation sessions of Section~\ref{sec:eval:sessions} are
  indicated with color. 
  The bottom shows the timing of various deployments with users.}
  This rich collection of over 150 artifacts mitigated issues in designing around
  shifting data and concerns.
  }
	\label{fig:teaser}
}

\vgtcinsertpkg

\begin{document}



\maketitle

\section{Introduction}
\label{sec:introduction}

When choosing whether to move forward with a design study there are several
questions a visualization expert should answer to ensure project
viability~\cite{Sedlmair2012}. Among those questions are: (1) whether real,
non-synthetic, data is available, and (2) whether the tasks that domain
experts will use the visualization for will persist long enough to complete
the study. Ensuring these points can help avoid problems arising in designing
for the wrong data assumptions or having the users lose interest before the
system is completed and evaluated. While in most cases it is prudent to avoid
these problems, we argue there are circumstances in which it is fruitful to
accept them.

In particular, there are scenarios where precisely what data to collect is
an open question. The answer would ideally be driven by what analysis needs to be
performed. This ``chicken and egg'' situation can dissuade both domain and
visualization experts from engaging. The domain expert does not want to
collect data with no plan for analysis. The visualization expert cannot act
without real data. Thus, an opportunity for visualization to inform the data
collection process is lost and the greater problem remains unsolved.

We observed this scenario in the domain of task-parallel computing. Parallel
systems are challenging to comprehend due to their complexity. Several factors
affect performance and correctness including program source code; the parallel
libraries used; the input to the program; and the architecture of the cluster on
which it is run. Understanding the emergent behavior in these systems is
necessary to optimize and debug them. While some areas of parallel computing
have a long history of data collection for performance analysis, the data
necessary to analyze a more recently prominent model---asynchronous many
tasks---is a relatively open area.

Recognizing the pitfalls of changing data and concerns, but also the potential
of using visualization to help drive the development of those data and
concerns, we proceeded with the design study. Although the pitfalls could have
derailed the project, we found other factors---such as the shared interest in
the data collection problem and the identification of key recurring abstract
structures---resulted in the creation of visualizations that were beneficial
even as things changed. From these experiences, we demonstrated how the
benefits of visualization are not only the design of the final solution, but
the integration of the visualization team and its effects on the overall
project development.

We describe our iterative design process and how we adapted the design study
methodology~\cite{Sedlmair2012} to the ``moving targets'' of our data and user
tasks. Our task analysis and abstraction were developed through multiple
rounds to account for evolving concerns.

Through this process, we developed our technology
probe~\cite{Hutchinson2003TechProbes}, Atria, a multi-view system for exploring
execution graphs. Unlike other execution graph visualizations, we display the
graph as an expression tree, evoking the main logical dependencies of the
computation while preserving additional edges on demand. Atria provides the
context of source code to the execution graph not only through linked
highlighting, but also through a default aggregation and de-cluttering scheme
based on line of code.

We further discuss how the changing concerns affected deployment with
implications for design, particularly in the case of Jupyter
notebooks~\cite{jupyter}. We augment our assessment of our design with
evaluation sessions and semi-structured interviews.

The major contributions of our study are as follows:
\begin{itemize}
  \itemsep=-0.5ex
  \item A task analysis for execution graphs
    (Section~\ref{sec:taskanalyses}) and how we iterated on that
    analysis under changing conditions,
  \item The design of Atria, a interactive visual tool for analyzing task
    execution graphs (Section~\ref{sec:visualizations}),
  \item A discussion of the design adaptations and the
    difficulties of incorporating visualization within the
    Jupyter notebook environment (Section~\ref{sec:vis:deploy}), and
  \item Reflection on the project and recommendations for conducting design
    studies in the presence of evolving data and concerns
    (Sections~\ref{sec:phylanx:winnow}, ~\ref{sec:phylanx:cast}, and~\ref{sec:reflections}).
\end{itemize}

We discuss related work (Section~\ref{sec:relatedwork}) followed by necessary
background in task-parallel execution graphs  
(Section~\ref{sec:ATR}). We then discuss the organization of the project
(Section~\ref{sec:phylanx:organization}). We conclude in
Section~\ref{sec:conclusion}.

\section{Related Work}
\label{sec:relatedwork}

We discuss related work in design study methodology, \forreview{task
abstraction}, visualization of execution graphs, and tree visualization
techniques.


\textbf{Design Studies.}
We report on the initial phases of an ongoing
design study,
including multiple deployments of our system, Atria, as a technology
probe~\cite{Hutchinson2003TechProbes}. This process has enabled us to
collect rich, qualitative data. While informing future
iterations in our collaboration with domain experts,
reflections~\cite{Meyer2018Reflection} (Section~\ref{sec:reflections})
on these data already have meaningful
implications for the visualization community.

In the context of the nine-stage framework for design study
methodology~\cite{Sedlmair2012}, this work represents a full cycle, including
many sub-cycles, of each of the nine stages. We build upon previous
visualization and design study experience at the \textit{learn} stage to
inform careful, deliberate decisions at the \textit{winnow} stage, that we
discuss in Section~\ref{sec:phylanx:winnow}. We observed refinements at the
\textit{cast} stage over time, where deployments and conversations with domain
experts exposed deeper insights into the various roles that they play in
practice. At the \textit{discover} stage, our tool enabled insights for our
collaborators and ourselves, particularly with respect to horizontal movements
in the task-information space---Atria drove many discussions about what data
needed to be collected that were unlikely to have occurred without our probe's
involvement. The simple nature of the tool enabled relatively simple
\textit{design} refinements, as well as rapid \textit{implement} and
\textit{deploy} stages. \forreview{Throughout we use the term {\it
iteratively} to describe the repetition of design stages and {\it iteration}
to mean each new version of Atria presented to our collaborators, as marked in
green in Figure~\ref{fig:teaser}}.

In framing our contributions in terms of the nine-stage framework, it is
important to address considerations that it does not capture. We deployed
Atria in very diverse technical and practical contexts, creating a parallel,
multi-channel \forreview{collaboration environment like that described in
detail by Wood et al.~\cite{Wood2014}. We did not experience the
constraints described by Crisan et al.~\cite{Crisan2016}, but  our process also
benefited from thorough artifact generation, frequent communication, and staged
design.} Our focus in this early phase of a long-term
collaboration has been to elicit robust design requirements, rather than
deploying prototypes too early~\cite{McKenna2017DAF}. Instead, we report on
the use of a technology probe to build, intervene, and
evaluate~\cite{McCurdy2016ADR}, with Sedlmair et al.'s~\cite{Sedlmair2012}
\textit{reflect} and \textit{write} stages applying continuously. As Hinrichs
et al.~\cite{Hinrichs2016} show in digital humanities, we demonstrate value in
the visualization {\em process}.  Finally, we contend that, in some cases, it
may be beneficial to consider collaborations where a design study will have an
opportunity to impact how data is collected and how the initial data
abstraction is designed.

\forreview{
\textbf{Task Abstraction.} Several methods have been proposed for bridging low
level visualization tasks to more complex
goals~\cite{Brehmer2013,Rind2016,Lam2017,Zhang2019}. Brehmer and
Munzner~\cite{Brehmer2013} propose composition of low level tasks. Zhang et
al.~\cite{Zhang2019} demonstrate combining hierarchical task
analysis~\cite{Annett1971,Annett2000,Salmon2010} with visualization task abstractions. As the data
and concerns evolved, we did not observe stationary tasks at a low enough
level to yet apply these detailed methods meaningfully. However, recognizing the importance
of linking goals and tasks, we describe our task findings in terms of multiple
levels. The lower level tasks of high level goals often overlapped,
resulting in a lattice as described in Section~\ref{sec:taskanalyses}.
}

\textbf{Execution Graph Visualization.}
Node-link diagrams are prevalent in execution graph
visualization~\cite{Brinkmann2013Temanejo, Dokulil2017, Huynh2015DAGViz,
Reissmann2017GrainGraphs}. Dokulil and Katreniakova~\cite{Dokulil2017} remove
edges that can be reached by other paths, plotting their results with
\texttt{dot}~\cite{Gansner93atechnique,GraphViz}.
DAGViz~\cite{Huynh2015DAGViz} and Grain Graphs~\cite{Reissmann2017GrainGraphs}
use aggregation schemes to decrease the number of marks shown, taking
advantage of nesting structures inherent to fork-join models of parallelism,
but not present in general tasking models such as ours. Neither solution
is interactive. In contrast, our interactive visualization abstracts execution
graphs to a tree and aggregates based on the source code.

Trace data from tasking models has been visualized with Gantt charts.
Ravel~\cite{Isaacs2015SC} shows the edges in Charm++ traces~\cite{charm}, but
these are a subset of those in the full execution graph. Haugen et
al.~\cite{Haugen2015} show edges connected to a single task on demand. Pinto
et al.~\cite{Pinto2016} do not show edges. We do not have trace data and thus
Gantt charts are inappropriate for our use. 

\forreview{Parallel calling context trees (CCTs) describe caller-callee relationships and
are frequently visualized~\cite{Ahn2009STAT, Adamoli2010Trevis, Lin2010,
Nguyen2016VIPACT, VAtrees}. They differ from our execution graphs which are at
a finer-grain task level and include dependencies not captured by CCTs.} For a
survey of visualizations across parallel computing models, see Isaacs et
al.~\cite{Isaacs2014STAR}.

\textbf{Tree Visualization.}
Visualizing a graph as a tree is an established
practice~\cite{Hao2000,Eklund2002,JankunKelly2003,Nobre2019Juniper,Nobre2019Lineage}
to reduce complexity and improve readability, especially when the tree has a
semantic meaning. In our case, we visualize the edges relating to how the
computation is expressed in code---its {\em expression tree}. 

To reduce clutter, we collapse subtrees and encode them as triangles,
like SpaceTree~\cite{Plaisant2002}, but our collapsing strategy is based
on meaning in the source code rather than screen space. Many
techniques~\cite{Lamping1995,Card2002,Plaisant2002,Munzner2003,Tominski2006}
exist for scalable hierarchy visualization. As part of the strategy of handling
evolving data and tasks, we aim to ``[satisfy] rather than
optimize''~\cite{Sedlmair2012} our depiction, but could apply these techniques
once necessary. For a survey of tree techniques, see Schulz~\cite{Schulz2011}.



\section{Task Parallel Programs and Execution Graphs}
\label{sec:ATR}

Parallel and distributed programs utilize vast computational resources to
produce results that are often too time-consuming, if not infeasible, on a
single processor. Achieving these time benefits often requires careful
consideration of performance on the part of the programmers. Understanding
observed performance is difficult because of the complexity of the emergent
behavior stemming from the source code and the systems on which they run. The
systems include not only the hardware, but the {\em runtime} which dictates
the execution of the program and the environment in which it runs.

Performance is affected by several factors including adaptive scheduling
policies that are difficult to predict. Intermediate structures may be created
by the runtime that are not apparent to the programmers. Even known structures
may be hard to reason about, given the dynamism in the system.

Asynchronous tasking runtimes (ATRs) are a class of parallel runtimes that
have been gaining interest for their potential to increase resource
utilization and overcome performance bottlenecks inherent to other paradigms.
However, due to their more recent prominence, support for performance analysis
of ATRs is still developing.

An {\em Asynchronous Tasking Runtime} supports an {\em Asynchronous Many-Task}
(AMT) execution model. Typically, these models divide and encapsulate work
(computation) into units known as {\em tasks}. The runtime then schedules the
task for execution on one of its distributed resources. The flexibility to move
tasks between resources allows ATRs to take advantage of parallelism other
models may not.

\subsection{Execution Graphs}
\label{sec:executiongraphs}

Common to tasking models is the notion of an {\em execution graph}. In an
execution graph, each task is a node. The edges are dependencies between
tasks. A task cannot be executed until its dependencies are met. Tasks without
dependencies between each other may run concurrently. Runtime developers are
thus interested in these dependencies and their effect on scheduling
decisions.

Execution graphs may be recorded during program execution. To reduce
collection overhead, tasks of the same type (e.g., same function name) or with
the same provenance (e.g., same function name and sequence of function names
leading to the task) may be aggregated. We focus on the latter type of
execution graph data in this project.

Attribute data is collected for each node in the execution graph. Typically
for performance analysis, the number of times each task type was run ({\em
count}) and the total time spent executing instances of that task are recorded.
During our project, attribute data was augmented to also collect information
about what mode the task was executed (see Section~\ref{sec:phylanx}) and
relation to line of source code.

\subsection{Performance Data and Analysis in ATRs}
\label{sec:atrperf}

There are many ATRs under active
development~\cite{hpx,charm,ocr,legion,parsec,uintah}, but no standardized
expectation for what performance-related data is collected. Existing parallel
performance tools like TAU~\cite{TAU} and Score-P~\cite{scorep} can collect
general performance data such as low-level profiles and traces, but often do
not support ATR-specific data such as execution graphs. Existing execution
graph work has often been specific to the ATR. Exactly what data could and
should be collected when analyzing an ATR is an open area of research.

\section{The Phylanx Project}
\label{sec:phylanx}

We conducted this design study as part of a visualization initiative in the
Phylanx Project. First, we provide technical background of the
Phylanx~\cite{Tohid2018} system necessary to understand our resulting
visualization. We then discuss the organization of the project itself, how it
led us to accept winnowing pitfalls, and the roles we cast in the design
study.

\subsection{Technical Overview of Phylanx}
\label{sec:phylanx:system}

Phylanx is an actively-developed system for performing array computations in a
distributed fashion. Its purpose is to provide the advantages of distributed
resources (faster time-to-solution and the ability to scale beyond
single-machine memory limitations) to data scientists while allowing them to
use the tools with which they are familiar.

One such tool is Python. Phylanx has a front-end which allows data scientists
to mark which of their Python functions they want run distributedly. Phylanx
then translates the array operations in the Python code into HPX.
HPX~\cite{hpx, hpxgit} is a C++ standard library and asynchronous tasking
runtime.

The average end users need not be concerned with how Phylanx transforms their
code. However, power users interested in performance and the developers of the
Phylanx system are.

Phylanx first translates the code into an intermediate representation in a
domain-specific, functional language called PhySL. The function calls, control
flow operations, data operations, and blocks found in the PhySL representation
are referred to as {\em primitives}. These primitives are translated to tasks
in the HPX runtime. The dependencies of each primitive are the arguments it
needs to execute (which may be other primitives), data access operations, or
any other constraints on variables the primitive uses.  The dependencies and
primitives form the execution graph which is run by HPX.

HPX can schedule any instance of a primitive in one of two modes: {\em
synchronous} or {\em asynchronous}. A synchronous primitive is executed
immediately from the primitive that initially spawned it. An asynchronous
primitive is added to an internal work queue and may be executed on a
different processor at some later time. Asynchronously scheduled primitives
give the runtime more flexibility but incur more overhead, so it is beneficial
to execute shorter primitives synchronously.

\subsection{Phylanx Project Organization}
\label{sec:phylanx:organization}

The Phylanx project comprises three teams, each located at a different
academic institution. The {\bf Runtime Team} develops the HPX and Phylanx
libraries. 
The {\bf Performance Analysis Team} develops instrumentation to collect
performance data and tools to improve performance. They also maintain the
nightly regression tests and reporting. The {\bf Visualization Team} develops
visual tools to aid in performance analysis and debugging. 
Additionally, a program manager (PM) for the project seeks out further
collaborations and develops data science applications using the Phylanx
system. \forreview{A list of} team members involved in the design
process and their roles \forreview{is available in the supplemental material}.


We discuss the inception and further organization of the project within the
framework of the design study methodology of Seldmair et
al.~\cite{Sedlmair2012}, specifically the {\em winnow} and {\em cast} phases.
We note which pitfalls were accepted and what other aspects of the project
helped mitigate the negative affects of those pitfalls.

\subsubsection{Accepting Winnowing Pitfalls}
\label{sec:phylanx:winnow}

Prior to the official project start, the Performance Analysis PI and the
Visualization PI had several conversations regarding difficulties in analyzing
ATRs. The Visualization PI had faced the issue of traditional data collection
being insufficient to analyze ATRs~\cite{Isaacs2015SC}. The Performance
Analysis PI expressed difficulty in making sense of the 
data that could be collected. He noted there was little point in spending
development resources and overhead on data that could not be analyzed.

These two coupled issues, (1) no data to analyze and (2) no analysis with which
to use to the data, present a ``chicken and egg'' barrier to improving
understanding and performance of ATRs. The two PIs view determining
what data to collect a research goal of the project.

The data being an evolving target of research, along with the development of
Phylanx itself and its changing concerns, means the project runs afoul of two
of Sedlmair et al.'s winnowing pitfalls:

{\bf PF-4: No Real Data Available (Yet).} During the project, the
structure of the data and the format of the data have been evolving. Other
potential sources of data are not yet instrumented. It is difficult to arrive
at a final visual solution without finalized data.

{\bf PF-10: No Real/Important/Recurring Task.} The fact that the data is in
flux means tasks involving that data are also in flux. Furthermore, as Phylanx
is developing rapidly, the concerns of the team members change over time,
affecting their higher-level goals.

The decision to proceed despite these pitfalls was motivated by the desire to
solve the larger problem of performance optimization and analysis for ATRs.
We view working with preliminary and in-flux data as a stepping stone to
achieving the ``data behind the data''---the data that can only be envisioned
with knowledge gained from exploring what we already know.

There are several factors that help with the continuing success of the
project, despite the pitfalls:

{\bf Identification and availability of meaningful preliminary data.} Though
the data collection is its own area of research, the PIs foresaw the
importance of the execution graph based on prior work and could confidently
predict it would continue to be useful to understand. We hypothesized that
some tasks would therefore remain stable (see Section~\ref{sec:taskanalyses}).
Furthermore, design did not begin until a preliminary dataset could be generated.


{\bf Strong interpersonal relationships.} The cohesion of the three groups
facilitated adaptation to new data. Teams were quick to clarify or explain
changes in format and to react to requests to change. The trust among the
teams allowed the Visualization Team to plan for future functionality with
relatively low risk.

{\bf Overarching goal of the project did not change.} The high level
goals of performance analysis and optimization, along with the goal of
discovering what data to collect, remained the same, though strategies
employed by the users changed. Thus, high-level goals that are
aided by visualization, such as understanding the execution, remained fixed.

{\bf Visualization considered a deliverable by entire project.} All project
teams recognize the visualization component as an outcome. Progress on the
visualization is reported at the weekly full-project teleconference and
included in all reports. The success of the project includes the success of
the visualizations.

The incorporation of visualization as a project-wide outcome underscores the
continuing approval and enthusiasm communicated by project gatekeepers,
\forreview{placing them in the High Power-High Interest quadrant of the matrix proposed
by Crisan et al.~\cite{Crisan2016}}. Team members were not only authorized to
spend time on the visualization, but encouraged to do so. We further discuss
project roles below.

\subsubsection{Casting Roles: Gatekeepers, Analysts, Experts}
\label{sec:phylanx:cast}

The Runtime and Performance Analysis PIs, project manager, and program manager
all serve in the gatekeeper role, with the Runtime PI and project manager
being the most central in allocating time with front-line analysts. One
student was identified as a front-line analyst early in the project. As the
project evolved, several other students with differing concerns (See
Section~\ref{sec:evaluation}) were cast in the role.

The gatekeepers also acted as front-line analysts. The PIs had similar
technical goals as the students. The project and program managers were more
representative of a second goal---communicating the project to outsiders. All
gave feedback regarding designs throughout the project.

An interesting facet of the project is that almost every person is a form of
tool builder. Sedlmair et al. noted the pitfall of mistaking fellow tool
builders for front-line analysts. Here they are both because a major goal of
the visualization is to help the tool builders in building their tools. 
Their role as tool builders furhter helped them accept working with an
in-development visualization (See Section~\ref{sec:reflections}).

Some studies have found success in blurring the boundaries between domain and
visualization experts~\cite{Wood2014}. Our project naturally maintained them,
further avoiding the pitfalls of working with fellow tool
builders~\cite{Bigelow2017}. We found communicating with mock ups and screen shots
was sufficient---users did not need to learn the language of visualization.
Furthermore, as the other teams trusted in the visualization expertise of the
designers, they accepted change in the design over time.

\section{Task analyses}
\label{sec:taskanalyses}

\begin{figure*}[htbp]
  \centering
  \includegraphics[width=\linewidth]{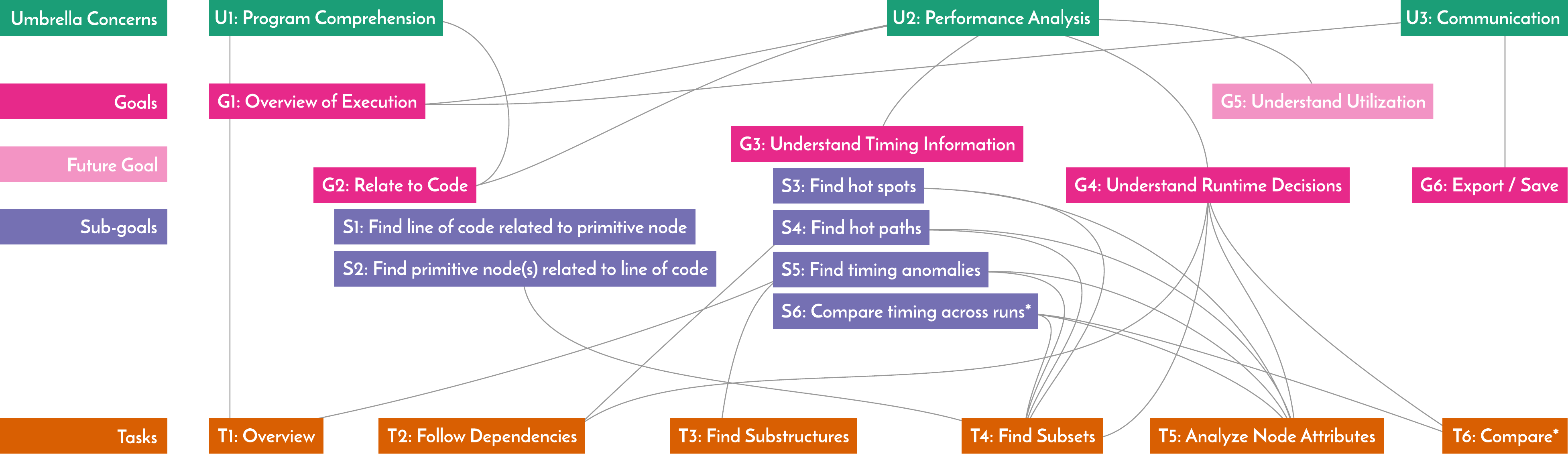}
  \vspace{-2ex}
  \caption{A goal-to-task lattice, showing the relationships between high-level
  umbrella concerns (U1 - U3); more specific goals (G1 - G6) and sub-goals
  (S1 - S6); and low-level tasks (T1 - T6) that directly inform the design of a
  visualization interface. The comparison sub-goal and task were added as data
  and concerns evolved. \forreview{G5 was identified as a future goal based on project
  priorities in collecting and analyzing the data.}
  \label{fig:hierarchy}
  }
\end{figure*}

We had three objectives in designing our visualization. We wanted to (1)
support the analysis needs of our collaborators, (2) refine data collection
and analysis for tasking models, and (3) prepare for future needs given the
refined data collection and the progress of the Phylanx project. Through our
multi-year collaboration, we assessed needs through general project meetings,
focused visualization and performance analysis meetings, and informal
interviews. From these, we developed a goal-to-task lattice (
Fig.~\ref{fig:hierarchy}), which we updated as needs shifted. We elaborate on
this process and present the lattice below.

The project had weekly status meetings where all teams gave updates on the
progress of individual components. Emerging problems were briefly discussed,
but scheduled for another meeting if necessary. There was an optional meeting
slot to discuss performance analysis and visualization specifically when
requested. We wrote notes from both these meetings, including subjects not
directly related to the visualization. We also had face-to-face meetings twice
a year, once at another team's site and another at a conference in high
performance computing.

Through the present, we created 152 note files with a mean 2800 characters per
file. Some contractual information prevents us from releasing the complete
audit trail~\cite{Carcary2009Audit} at this time, but anonymized summaries of
our task analyses, with relationships between specific note files and the
goal-to-task lattice are included as supplemental material. The project
manager also compiled regular notes from the perspective of the Runtime team
which augmented our understanding of the full project status and aided in our
planning.

Tasks regarding the execution graph were derived from the note files by two
authors independently who then developed a
lattice spanning from high level goals to low level tasks using affinity
diagramming. We classify these as {\em umbrella concerns} (U1 - U3), {\em goals}
(G1 - G6), {\em sub-goals} (S1 - S6), and {\em tasks} (T1 - T6).

\subsection{Umbrella Concerns}
\label{sec:umbrella}

We use the term {\em umbrella concerns} to describe the major classes of goals
we found our users had with respect to visual analysis. Some goals fell under
multiple umbrella concerns.

{\bf U1. Program Comprehension.} Our collaborators want to understand what
happened when the program was executed. Many were working on a specific piece
of the Phylanx pipeline and did not have a concrete mental model of how the
translation from code to execution graph took place, nor a sense of the
intermediate PhySL representation. Although previous work found that computing
researchers may consult graphs to debug their mental model~\cite{Devkota2018},
we found some of our collaborators wanted to build their mental model. This is
often a first step to devising new strategies, debugging, or performance
analysis.

{\bf U2. Performance Analysis.} An impetus for moving to tasking runtimes is
the potential for high performance---decreasing the time to solutions and/or
making previously infeasible computations feasible. Thus, understanding and
improving the performance of a given Phylanx application or the system itself
was a driving concern.

{\bf U3. Communication.} Our collaborators wanted to create figures to help
explain their own research in publications. The project and program managers
were interested in explaining to potential users how the Phylanx system works.
Such users often already have a background in parallel computing and thus can
interpret the visualization when presented by someone from the team.

\subsection{Goal-Task Lattice}
\label{sec:goaltask}

We identified six goals relating to the execution graph and our umbrella
concerns, some of which could be divided into smaller sub-goals. We discuss
each goal and relate it to low-level tasks. We then summarize the tasks pulled
from our goals.

{\bf G1. Overview of Execution.} All three umbrella concerns wanted some sort
of overview of what happened during the execution, in particular, the size and
shape of the execution graph and how many times each node was executed. This
goal can be divided into tasks of gaining a graph overview (T1), following
dependencies (T2), and finding substructures (T3). For example, our
collaborators explained that the visualization should allow them to understand
if something was called recursively. This can be done by following a cycle of
dependencies in the aggregated execution graph.

{\bf G2. Relate to Code.} While the execution graph describes how the runtime
executes the program, our collaborators cannot directly change the graph
itself, only the associated source. Thus, they want to know the relationship
between the code and the graph for both program comprehension and performance
analysis concerns. We divide this into two sub-goals: (1) finding the line of
code related to a node in the graph and (2) finding the nodes in the graph
related to a line of code. The latter we categorize as a task of finding a
subset of nodes (T4).

{\bf G3. Understanding Timing Information.} Central to the Performance
Analysis concern is data recorded about time spent executing each node. Of
particular interest is finding parts of the execution that took a long time or
behaved in an unexpected way, leading us to identify sub-goals of: (1) finding
hot spots, (2) finding hot paths, and (3) finding timing anomalies. Hot spots
are nodes that executed for a long time. Hot paths are sequences of such
nodes.

Later in the project, as our collaborators progressed from the initial
development of their applications to performance optimization, a fourth
sub-goal, (4) comparing performance between runs, was discovered. We added it
when we revisited our goal-task lattice.

All of these sub-goals require finding a subset of interesting nodes (T4) and
analyzing attribute data of those nodes (T5). The hot paths sub-goal also
requires following dependencies (T2). The timing anomalies sub-goal may
further involve identifying substructures in the graph (T3) and understanding
an overview (T1). The comparison sub-goal requires comparing attribute data
(T6).

{\bf G4. Understand Runtime Decisions.} A key feature of tasking runtimes is
built-in support for adaptively altering execution based on runtime data to
improve performance. Our collaborators want to know what choices were
made and the effect on performance, making this a Performance
concern. An example of this goal is the choice of execution mode as described in
Section~\ref{sec:phylanx:system}. Similar to G3, understanding runtime
decisions is aided by finding a subset of interesting nodes (T4) and analyzing
node attribute data (about runtime parameters) (T5). As decisions are related
to dependencies, following dependencies (T2) is another task. Also like G3,
we updated the tasks for this goal with comparison (T6) as the objectives of
our collaborators shifted.

{\bf G5. Understand Utilization.} The primitives represented by the execution
graph as nodes must be scheduled to run on computing resources. Researchers
are interested in maximizing the utilization of those resources---spending
less time idling and more time doing useful work. Thus, this was another
Performance concern. \forreview{However, neither examining this data nor the
capability to associate utilization data with the execution graph was a
development priority for our collaborators over the other goals.} We thus
included it in our goal-task lattice as a possible future node, with
references to notes on the matter so we may look back on them should
utilization become a more pressing concern.

{\bf G6. Export/Save.} Supporting the Communication concern, our collaborators
requested a mechanism for exporting and saving the visualization.

From the goals, we collect six low-level tasks. We list them here followed by
their relationship to the task taxonomy for graphs of Lee et
al.~\cite{Lee2006}.

\begin{itemize}
  \itemsep=-0.5ex
  \item[T1.] Overview (4.4 Overview)
  \item[T2.] Follow Dependencies (4.1.1 Adjacency, 4.3.1 Follow Path)
  \item[T3.] Find Substructures (5 Higher Level Tasks)
  \item[T4.] Find Subsets (4.2.1 Node Attribute Tasks)
  \item[T5.] Analyze Node Attributes (4.2.1 Node Attribute Tasks)
  \item[T6.] Compare (4.2.1 Node Attribute Tasks, 5 Higher Level Tasks)
\end{itemize}

The presence of topology-based (finding adjacencies) and browsing (follow
path) tasks when understanding dependencies motivates our use of visual
representations that explicitly encode edges.

While there are several node attribute tasks, we note that we have relatively
few attributes---timing data and mode of execution. This motivates our design
decision to use on-node encoding, as it is easily understood by most
users~\cite{Nobre2019Star}.

\begin{figure*}[htbp]
  \centering
  \includegraphics[width=\linewidth]{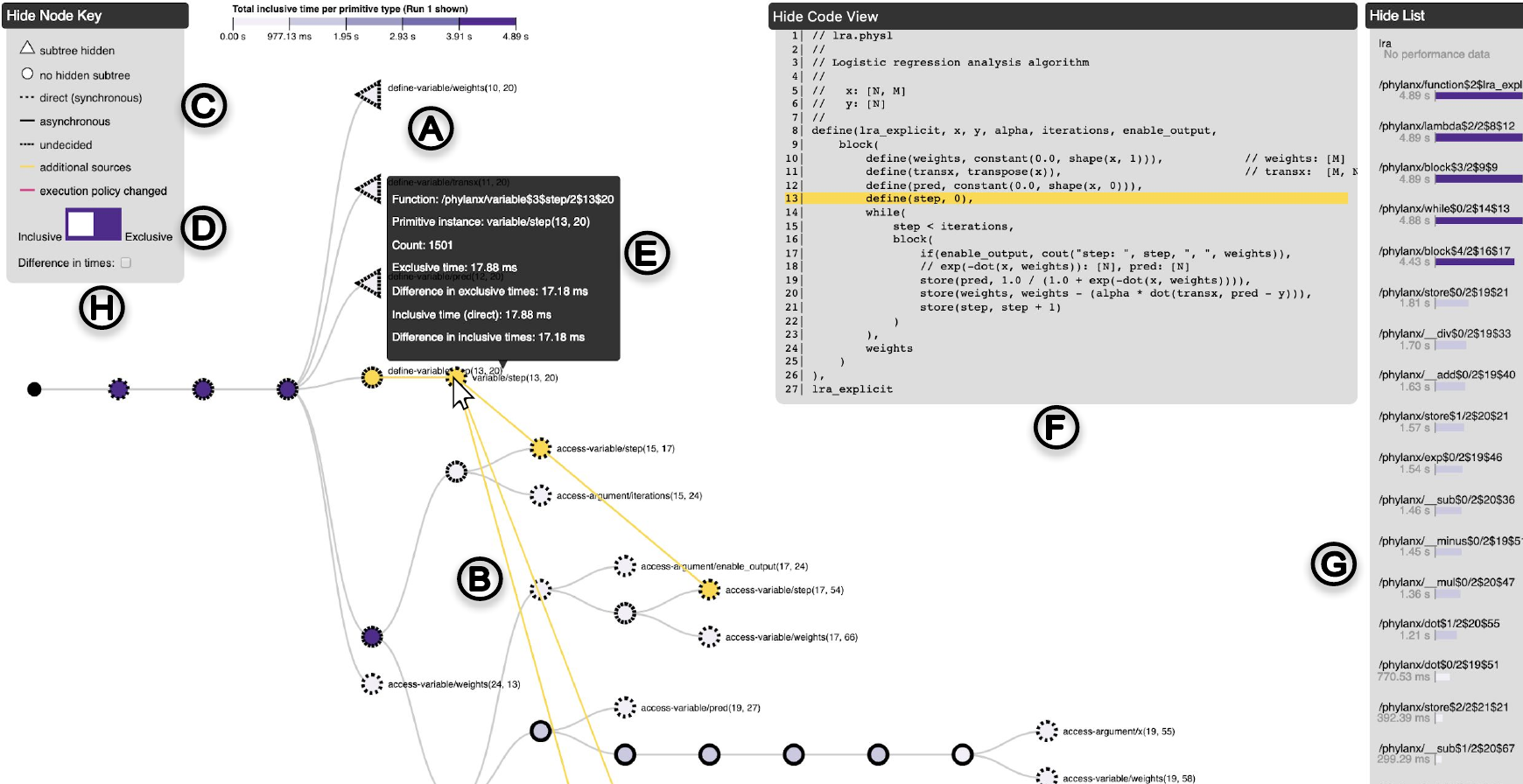}
  \vspace{-2ex}
  \caption{The design of Atria. The main view represents the expression tree
  contained in the execution graph. (A) Triangles represent collapsed
  subtrees. (B) Elided links are shown on hover. (C) Fill color and border
  style encode time and execution mode respectively. (D) Users can toggle
  between showing inclusive and exclusive time. (E) Tooltips provide details
  on hover. (F) Code view with linked line of code highlighting. (G)
  \forreview{Primitives}
  listed by execution time. (H) If multiple runs are available, comparative
  mode may be enabled.
  \label{fig:atria}
  }
\end{figure*}

\subsubsection{Evolution of the Goal-Task Lattice}
\label{sec:goaltaskevo}

We remark that our task analysis remained stable through multiple
\forreview{revisions}.  Later notes tended to reinforce goals and sub-goals
already in the lattice.  This may be due to the central need for comprehension
of the execution as a starting point for any other goal. We hypothesize this
relative stability over time contributed to the success of the visualization,
despite the evolution of the data and the shift in focus towards comparison.

\section{Visualization Design}
\label{sec:visualizations}

Atria (Fig.~\ref{fig:atria}), was designed and developed
iteratively as data became available. We describe our design choices and
explain how the evolving data, tasks, and environment influenced our design
decisions.

The central view of Atria is the execution graph, visualized as a node-link
tree. We explain this choice along with the choice of attribute encodings. We
then describe the auxiliary linked views.

Throughout its development, Atria has served several purposes: (1) an initial
validity check on data generated, (2) a visual tool supporting our
collaborators in their evolving tasks (Section~\ref{sec:taskanalyses}), and
(3) a platform for hypothesizing about what new data to collect to help with
the analysis. In support of these concerns, deployment of a working version
was a priority. Matching the evolution of project concerns, we strongly
embraced the advice~\cite{Sedlmair2012} of satisfying needs rather than
optimizing them. We describe the effect of these deployments on design,
including significant changes for Jupyter Notebooks
(Section~\ref{sec:vis:deploy}).

\subsection{Execution Graph}
\label{sec:vis:tree}

An execution graph is a directed acyclic graph of tasks describing the
dependencies that must be met before any task ({\em primitive}) can be
executed (Section~\ref{sec:phylanx}). Rather than show all edges in the graph,
we display a subset of the edges and lay out the graph as a tree.
Specifically, we represent the execution graph as an {\em expression tree}.

\begin{figure}[htbp]
\centering
\includegraphics[width=0.8\linewidth]{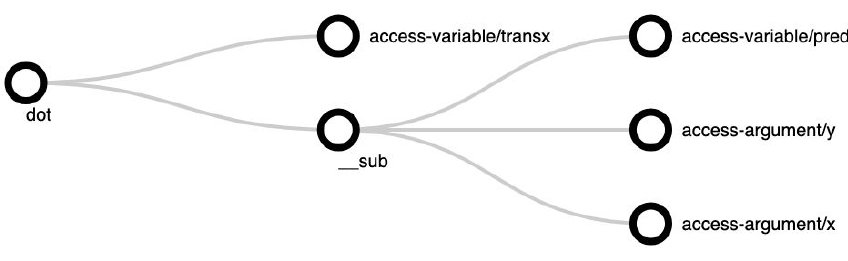}
\caption{The expression tree of $\texttt{transx} \cdot (\texttt{pred} - \texttt{y} - \texttt{x})$.
\label{fig:extree}
  }
\end{figure}

In an expression tree, each node is an operation and its children are its
operands. In Atria's graph view, each node is a primitive, which may be a
simple or complex operation, and each child is an operand to that primitive as
described by the PhySL intermediate representation. Fig.~\ref{fig:extree}
shows a small example. We chose to prioritize expression tree links because of
their relation to the PhySL code and to descriptions of the Phylanx model we
had gathered from discussions with collaborators and their presentations.

Our collaborators' interest in the expression tree abstraction drove the
evolution of the data collection. First, we collected only expression tree
data. We created three interactive tree visualizations using icicle plots,
node-link diagrams, and indented trees. Each allowed collapsing of sub-trees
into single marks (triangles in our node-link tree). Within a few months, it
became clear from viewing only the expression tree that there are cases where
the execution graph is needed for analysis.


We created mock ups showing a full-graph node-link diagram as well as options
for augmenting the tree visualizations with the extra edges. Our collaborators
uniformly preferred the tree layout. We then decided to focus on the node-link
representation (using D3's~\cite{Bostock2011d3js}
Reingold-Tilford~\cite{Reingold1981} layout) for the tree because that early
visualization received the most use; node-link diagrams are already prevalent
in the computing space~\cite{Isaacs2019}; and studies have shown the utility
of node-link representations for path following
tasks~\cite{Ghoniem2004,Keller2006} \forreview{i.e., T2}. 

Mindful to avoid premature design commitment, we revisited the choice of tree
representation later with a collaborator not involved earlier. He strongly
preferred the node-link tree, saying ``Whenever we were learning algorithms or
something like that, we would draw it like that. It's more comfortable because
we're more used to it and we can more easily see what's going on...Although
the one in class might be drawn top down though." We chose the horizontal
aspect ratio because most displays have more horizontal space. 

\forreview{This familiarity with node-link diagrams can further help users in
identifying substructures (T3). The trade off is that node-link diagrams are
not compact. Users had to zoom or scroll to gain an overview (T1).}

{\bf Elided Structure and Interaction.}
By showing only the expression tree as dictated by the PhySL, we have removed
two classes of edges: (1) dependencies between multiple accesses of the
same variable, and (2) dependencies between multiple uses of the same
function call. These are not usually needed for our users' goals. To
support a more detailed analysis if necessary, we show them in a node-centric
manner on demand. When a user hovers over a node, the edges are overlaid in
yellow to be non-obtrusive.

Some sub-structures in the tree are common but rarely of interest to our
collaborators. To de-clutter the visualization, we use two strategies. First,
we automatically collapse sub-trees for a known set of ``uninteresting''
primitives. Second, for the on-demand links, we omit edges between library
functions (e.g., \texttt{add}) as these do not communicate the structure of
the application, but lower level information about the runtime that our
collaborators do not expect to be of use.


{\bf Node Attribute Encodings.}
As timing and runtime decisions tie directly into our collaborators' goals
(G3, G4), \forreview{in particular task (T5),} we encode timing and execution
mode on-node. Execution time is encoded in the node's fill
saturation\forreview{, allowing users to find groups of nodes with similar
timing in context (T3, T4).} Execution mode is shown in the border's line style.  The exact
time and name of execution mode, as well as other attributes such as the
execution count, are provided in a tooltip on node hover \forreview{further
supporting T5}.

Users can switch between two concepts of execution time.  Inclusive
time is the wall-clock time taken for the primitive to execute.  Exclusive
time subtracts from the primitive any time that can be attributed to waiting
on its children. Our collaborators are interested in both.


\subsection{Auxiliary Views}
\label{sec:vis:views}

A collapsible linked view displayed the PhySL code to support G2, relating to
code. Sub-goals S1 and S2 are implemented as linked highlighting. The code
view auto-scrolls on node hover. We experimented with showing the related C++
or Python, but PhySL was preferred.

To support S3, finding hot spots, we have a collapsible list view which shows
the tree primitives from most time-consuming to least with colored bars
matching the on-node time encoding. The view is similar in design to that of
Intel's VTune, which is used by several of our collaborators. VTune works at a
lower-level of abstraction and cannot list by primitive.


\subsection{Designing for Comparison}
\label{sec:vis:compare}

During our design study, some of our collaborators began exploring the effect
of adaptive policies that change execution modes at runtime. They make changes
to these policies between runs and wish to compare the results. They reported
opening multiple Atria instances. In response we added a comparison mode.

Discussions after we proposed a comparison view indicated that our collaborators
only compare two runs at a time, allowing us to calculate a simple derived
value. For timing comparison, we change the node fill from execution time to
execution time {\em difference} using a diverging color scale. For execution
mode comparison, we highlight (magenta) the borders of primitives that were
executed differently between runs but keep the line-style encoding of the
first dataset. These encoding changes support the discovery of interest
subsets of nodes (T4) and their comparison (T6).

When only policies are changed, the structure of the tree does not change.
However, when the application code or Phylanx changes, the tree structure will
change. As Phylanx is under active development, we observed small topology
changes every few weeks. When we observe nodes that are not present in both
trees and thus cannot be compared, we draw them with lowered opacity, similar to
the approach employed by Campello et al.~\cite{Campello2014}. So far topology
comparison has not been a focus of our collaborators.

\subsection{Design Changes for Deployment}
\label{sec:vis:deploy}

Our primary Atria audience uses a web-based deployment. \forreview{As the data
collection, output, and use scenarios evolved, we created several design
variants, resulting in multiple similar deployments, described in the
supplemental material.}

\forreview{Output and collection changes, made by the Runtime team, were
driven by visualization goals, specifically: (1) integration of Atria with
automated nightly regression tests and (2) a full application-to-analysis
demonstration in Jupyter, requested by team members with external
communication goals (U3). We discuss changes for the latter below.}

\subsubsection{Atria in Jupyter}
\label{sec:jupyter}

Jupyter Notebook is an interactive coding environment supporting literate
programming. Users enter code into input cells that can be run (and re-run) to
produce output cells. Variables persist through multiple input cells.  Jupyter
Notebooks are one of Phylanx's front-ends, available through Docker
containers. The front-end is important to the project due to the ease and
portability of container installation combined with the prevalence of Jupyter
in the data science community. 

\begin{figure}[htbp]
  \centering
  \includegraphics[width=\linewidth]{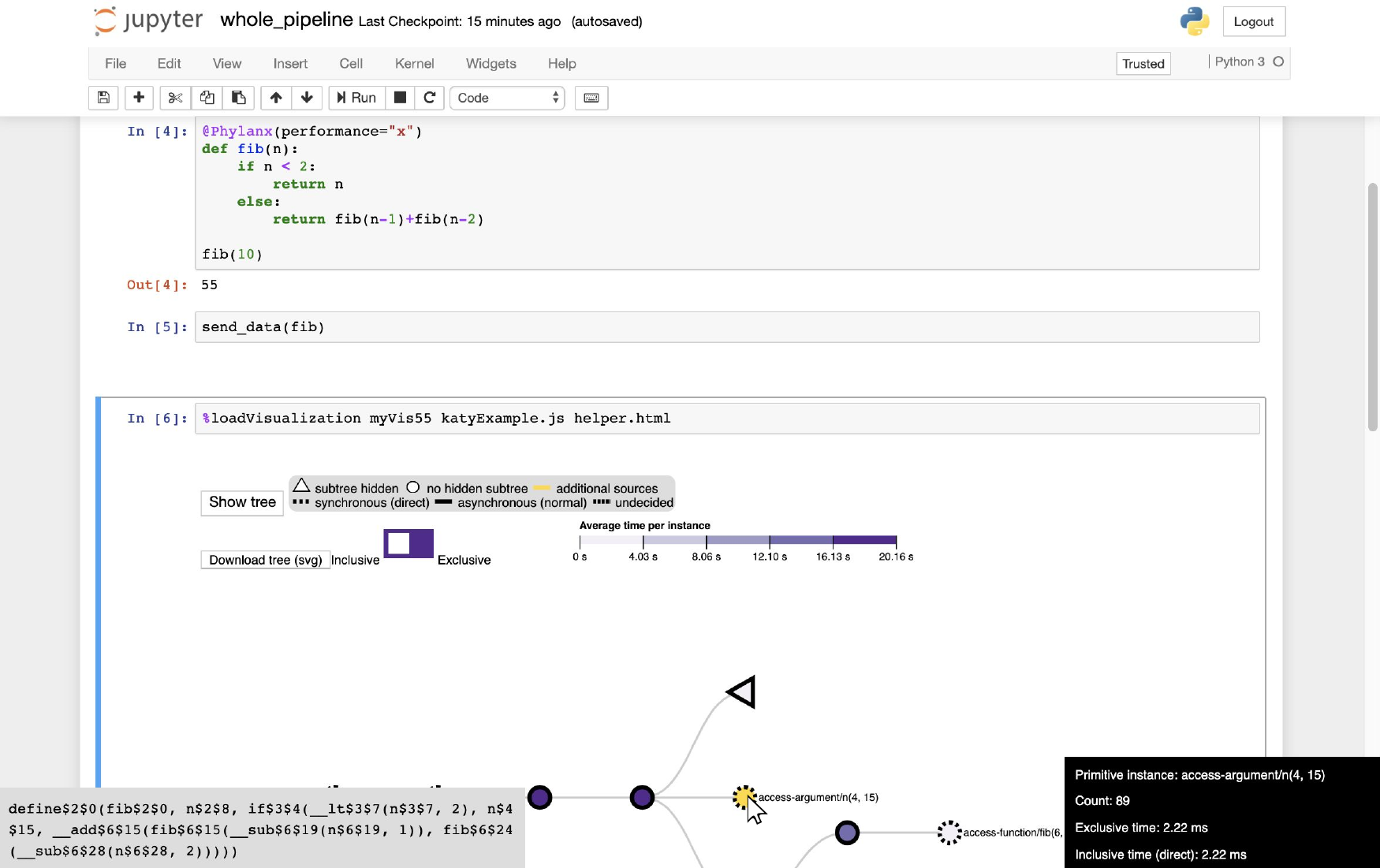}
  \vspace{-2ex}
  \caption{Atria in Jupyter Notebook. Cell 4 is Python code that uses the Phylanx library. 
  The newly-generated performance data and tree are passed in Cell 5. 
  Atria is loaded and displayed in Cell 6, with the generated PhySL shown in the bottom left 
  and the details of the hovered node shown in the bottom right. 
  \label{fig:jupyter}
  }
\end{figure}

The project and program managers give Phylanx demonstrations through this
front-end and thus wanted Atria integration to help explain the programming
model. Once we had an initial Jupyter pipeline in place, we found additional
users who wanted to test out small code snippets and see the effects. An
example notebook is shown in Fig.~\ref{fig:jupyter}. The export/save (G6)
functionality was prioritized as users wanted to further share results.

The Jupyter Notebook interface imposed an additional space constraint on
Atria, decreasing the width to $\approx 60\%$ of the browser. Normally our
users can devote an entire display to the visualization. We modified Atria's
layout of auxiliary views to prioritize visibility of the graph. Tool tip data
was moved to a fixed position in the bottom right so it did not obscure the
graph or legend.

We decreased the size of the code view and placed it in a floating window in
the bottom left. It shows three lines of code, which we determined was enough
context for our users during formative evaluation. Users can output the full
PhySL to a separate Jupyter cell, which was done during demonstration. The
Jupyter interface itself thus acted as another (full) code view, available via
scrolling.

Jupyter is hosted in a web environment with its own structure, styling, and
handling of Javascript. This posed technical challenges in embedding
our Javascript visualization in a cell in a maintainable manner. We view
streamlining of this process as an avenue for future work.

\section{Evaluation}
\label{sec:evaluation}

We evaluate Atria and its inclusion in the Phylanx project through case
studies gathered during deployment and evaluation sessions.
\forreview{Additional figures describing the evaluation and a video showing
the case study of Section~\ref{sec:eval:deploy:bibek} are included as
supplementary material.}

\subsection{Deployment Case Studies}
\label{sec:eval:deploy}

As described in Section~\ref{sec:vis:deploy}, we prioritized deploying
versions to our collaborators, creating several variants during the project.
Additionally, the program manager created a variant for his workflow at a
secure facility.
Data collection and design streamlining done for earlier deployments made this
last deployment possible.

\forreview{We polled our collaborators for their non-evaluative uses of Atria
every few months. R3 consistently reported using Atria as described below. One
other student reported using it actively when they were working on a
particular algorithm, but has since changed objectives and does not presently
use it. The program manager reported using it sporadically to explain the
project to others. In the evaluation sessions
(Section~\ref{sec:eval:sessions}), four participants report using it
minimally.}

We describe two case studies. The first shows how Atria is used
regularly in Phylanx development. The second describes how Atria was used in a
reactive situation to aid in reasoning and how the Atria development process
influenced the performance debugging process.

\subsubsection{Atria in Regular Use}
\label{sec:eval:deploy:bibek}

Our primary frontline analyst, R3, began using the deployed version of Atria
within a few months of the start of design \forreview{in January 2018. He reported using the
visualization on average once a week, more frequently when actively
debugging.}

\begin{figure}[tbp]
\centering
\includegraphics[width=1.0\linewidth]{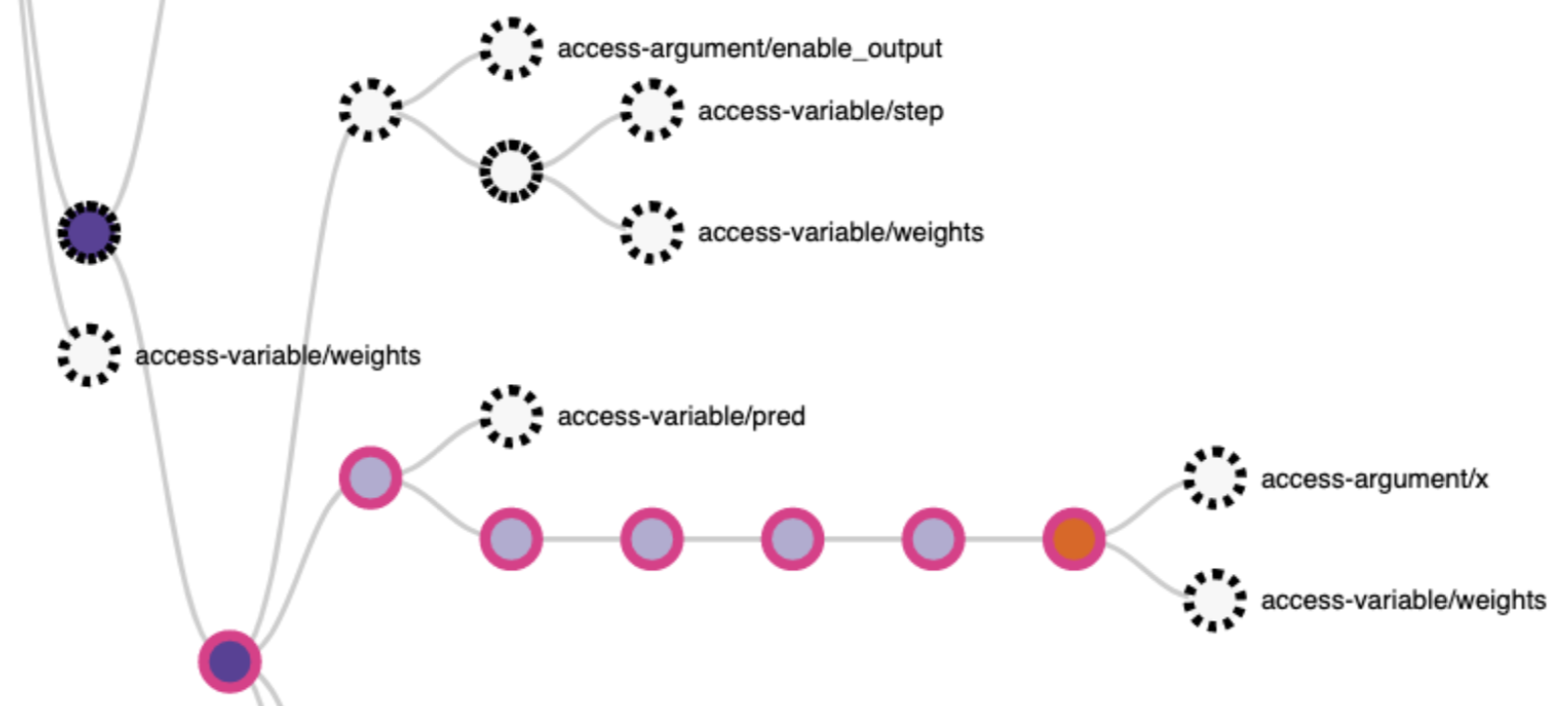}
\caption{\forreview{Comparison between two runs of the same application with different
policies. Pink-outlined nodes indicate a difference in execution mode between
two runs. The orange node ran slower after the policy change, but the net affect on
the parents was positive.}
\label{fig:compare}
  }
\end{figure}

He first runs the application he wishes to examine, generating the data used by
Atria. He copies the files to a local directory and opens Atria from the
command line. He considers the overall shape of the tree, noting that nodes
with similar depth may be candidates to run concurrently. Then, he considers a
particular primitive and its children to examine how the timing and execution
of the children may have affected the parent as shown in
Fig.~\ref{fig:compare}.

Using his gained intuition, R3 makes a change in the Phylanx policy. He
changes the thresholds that determine whether a primitive will be run
synchronously or asynchronously. He runs the program with the new policy and
collects data. Using Atria, he compares the two runs to see which of the
primitives changed their execution policies and whether that caused them to
run faster or slower. As the policy change is global and timing changes may
have non-local scheduling effects, he browses the entire tree. He uses his
findings to inform the next iteration of policy development.

When explaining his workflow to us, R3 said ``Also it's that I
want to be able to visualize it [the algorithm], just seeing it implants it in
my mind." He explained that he is a visual person and Atria makes it easier to
think about the problem.

\subsubsection{Investigation of Performance Regression}
\label{sec:eval:deploy:als}

A significant slowdown in Phylanx's alternating least squares (ALS)
application was discovered through nightly regression tests. 
The project manager (R2) suggested
using Atria to compare runs before and after the performance drop.

The regression tests ran with a large and a small dataset. Atria data was
collected only for the small run and showed no odd behavior, indicating
further examination of the larger run was required. As a test dataset, the
visualization team collected data using the older (pre-slowdown) code on a
different cluster. No performance difference was observed, indicating the
behavior was machine-specific.

\begin{figure}[tbp]
\centering
\includegraphics[width=1.0\linewidth]{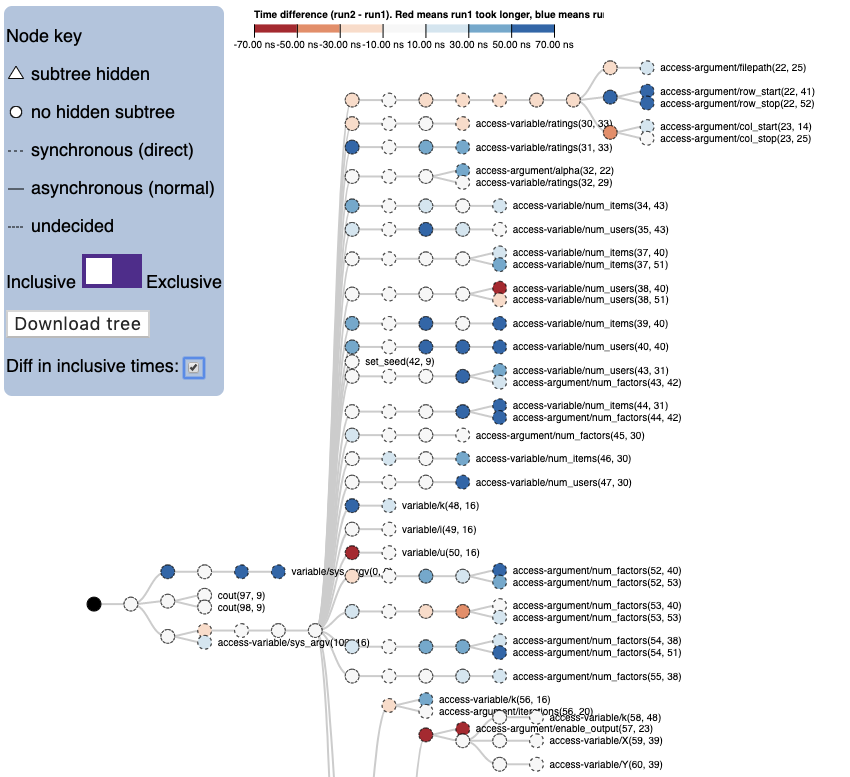}
\caption{\forreview{Comparison between two ALS runs, before and after a
  significant slowdown. As hypothesized by the Runtime PI, there would be a
  slight increase in execution time on average for the slower run (blue).}
\label{fig:alsregression}
  }
\end{figure}

The Performance Analysis PI (P1) then collected the larger run data on the
regression machine. He discovered the problem was due to a change at the HPX
level. He suggested it would therefore not be visible in Atria. The Runtime PI
(R1) hypothesized it would show up as a 10-30\% increase in all primitives on
average. We used Atria to compare the two versions
(Fig.~\ref{fig:alsregression}) and found R1's hypothesis to be correct.

Atria did not pinpoint the source of the problem, but was used to narrow the
space of possibilities and then confirm understanding of lower level effects
on the application. Furthermore, the involvement of Atria motivated deeper
examination of the problem and the data collection that led to discovery of
the root cause which was then fixed.






\subsection{Evaluation Sessions}
\label{sec:eval:sessions}

We conducted evaluation sessions of Atria with seven members of the Runtime
Team (R4--R10). R10 had no prior experience. Some had seen Atria briefly (R5,
R6, R8, R9), though R8 only remembered after completing the tasks and R9 had
only seen a picture. Two (R4, R7) had previous influence on design. Sessions
were conducted at the participant's workstation, with the exception of two
(R6, R7) which were done in a nearby meeting area.

We began our evaluation sessions with a demonstration and feature overview
using a small example. Participants could use Atria and ask questions.
We then asked users to perform a series of tasks on a dataset generated from a
Phylanx application with which they were not familiar. We followed up with a
semi-structured interview and de-briefing.

\subsubsection{Evaluation Sessions: Tasks}
\label{sec:eval:sessions:tasks}

We asked the following evaluation tasks, each marked with its
corresponding goal from Section~\ref{sec:taskanalyses}. Tasks \forreview{L1--L4 are for
a lone run.} Tasks C1--C2 are for comparative runs.

\vspace{1ex}

\forreview{L1:} Find a primitive that takes a lot of time. (G3)

\hspace{3ex}
How long does it take without its children? With? (G3)

\vspace{0.25ex}

\forreview{L2:} Find a primitive that is executed synchronously (G4)

\vspace{0.25ex}

\forreview{L3:} Find a primitive that is executed asynchronously (G4)

\vspace{0.25ex}

\forreview{L4:} Find a primitive that is repeated in the code (G1)

\vspace{0.25ex}

C1: Which run was slower? (G3),

\hspace{3ex}
*Why might it have been slower? (G1, G2, G3, G4)

\vspace{0.25ex}

C2: Find a primitive that changed execution mode. (G4)

\hspace{3ex}
Explain the change. (G4)

\vspace{1ex}

Five of the participants were able to complete the \forreview{L1} tasks
\forreview{within seconds}, doing a
visual search for the most saturated nodes and then reading the exact numbers
from the tool tip. R5 attempted to use the list view, but it wasn't yet
linked. They also required a reminder of the encoding before completion. 
R10 had difficulty and seemed to be still learning the
visualization. 

All participants were able to complete \forreview{L2 within seconds. However,
in L3, four the participants took tens of seconds.} We believe this was because there was only one
correct answer for L3 in the sample data. 
\forreview{L4} was also completed by all
participants, though two (R7, R10) asked for clarification.

In the comparison tasks, all participants answered which run took longer (C1),
with many verbally reasoning about the colors. However, in two cases (R4, R5),
the phrasing of the question accidentally included the follow-up hint. In
finding the changed execution (C2), all participants with the exception of R8
completed the task, though two asked that the encoding be re-explained. While
the other participants seemed to browse for a node, R8 flipped between the
comparison and non-comparison mode, appearing to search for line style changes.

C1* was a higher level analysis task that required performance analysis
background, thus we only asked it of participants who indicated they performed
such analysis in their duties (R5--R7). 
Each pointed out highly saturated nodes of the slower run's color as contributing
to the slowness and noted in particular a \texttt{store} primitive was among
them. R7 suggested that since the \texttt{store} took a lot of time, the
program might be memory-intensive. R5 noted he had outside knowledge---that
the \texttt{store} primitive had been modified recently---and concluded the
data might represent the performance change due to that modification.

Throughout the tasks, we noticed a few common themes in the interaction. In most
cases, participants appeared to use the encodings and interactions as intended.
A few verbalized their rationale. Several participants consulted the legend in
solving tasks (R6, R8, R10). Difficulties arose in discerning the line borders
(R4, R10) and in evaluating the last hovered node which remained yellow (R4, R5,
R7).

\subsubsection{Evaluation Sessions: Interviews}
\label{sec:eval:sessions:interviews}

We conducted semi-structured interviews with
participants, asking if there were any features they found useful and what
they would like the visualization do that it could not already. If
participants indicated they had used the visualization before, we asked them
how they used it.

Regarding utility, two participants said they didn't know whether the features
would be helpful or not (R6, R9). The remaining participants each listed
several components of Atria, but there was little consensus among them.
Repeated features included: access to timing data (P4, P5, P7), the linked
code view (P4, P5, P8), the comparison view (P4, P5, P9), and links between
dependencies (P5, P7, P8). Suggestions for improvement included
differentiating primitive types (e.g., variables, functions, control-flow)
(R6, R7) and more de-cluttering of the node-link tree (R6, R7).

Three participants (R4--R6) said they used a previous \forreview{deployment} to draw
figures for a paper~\cite{Tohid2018, Wagle2019} or report. R8 said they
used it to view the structure of codes they were not familiar with and see
timing data.

\subsubsection{Evaluation Sessions: Discussion}
\label{sec:eval:sessions:discussion}

In general, participants performed well on the evaluation session tasks with
indications that our encodings were used to complete them. Most evaluation tasks
were short as they were constrained by session length, but we consider them
necessary to establish basic usability and to engage participants for the
interviews.

\forreview{The least experienced participant, R10, struggled with several
tasks. R10 had just begun learning how to code. Our goal-task lattice focused
on expert analysis or communication led by an expert, which may explain these
observations.}

The participants who performed the high level task, hypothesizing why one run
was slower, showed a combination of reasoning including identifying hot spots
(G3) and relating knowledge about code (G2).

The sessions also revealed confusion in the execution mode encoding. It was a
recent change to support the new {\em undecided} mode. We attempted to
preserve the previous encoding of solid and dashed borders with undecided
being in between to match semantics. We are now reconsidering this choice. The
confusion reveals a design challenge due to the shift in what we could assume
about the data and a trade off with consistency with prior encodings.

We attempted to keep the interview question regarding useful features
unbiased. We first asked if there are any features that were useful,
stating that {\em no} and {\em I don't know} were helpful answers. Even still,
we suspect participants were predisposed to answer positively. Though
potentially biased, we were surprised by the variance in which features the
participants found potentially useful.

We hypothesize this variance is due to the differing concerns of the
participants. Some were starting to consider comparative performance analysis.
Some were focused on development of specific Phylanx features or example
applications.
The two participants who answered
they didn't know are focused on developing interfaces between existing
libraries and Phylanx. They are not working on the execution of the task graph
itself or on applications that decompose into the graph. The other
participants do but in different contexts, perhaps leading to their difference
in feature preferences.




\section{Reflection and Lessons Learned}
\label{sec:reflections}

We discussed earlier (Section~\ref{sec:phylanx:winnow}) the importance of the
respect the Phylanx teams have for each other and the positioning of the data
collection and visualization as goals of the project as a whole. These themes
carry through several of the other lessons we learned throughout the project
thus far. We discuss these lessons below.

{\bf When designing for a moving target, seeking to satisfy rather than
optimize is essential.} Recognizing that we were managing Sedlmair et
al.~\cite{Sedlmair2012} pitfall PF-10, {\em No Real/Important/Recurring Task},
we focused strongly on satisfying needs and deploying. We took a ``wait and
see'' approach with optimizing particular encoding choices and functionality,
not wanting to expend effort on features that might not last.  This mindset
also helped avoid pitfall PF-20, {\em Premature Design Commitment} as we
espoused the changing nature of the data and project.

Similarly, our rapid deployments often contained UI bugs. These primarily
decreased usability but did not change the meaning of the data---again
satisfying rather than optimizing. We believe these were accepted by our
collaborators because of the nature of active development throughout the whole
project. The Visualization team reported runtime bugs to the Runtime team
(opening tickets on Github), so the Runtime team naturally reported bugs with
the visualization.

{\bf Task analysis and long-term corpus of notes help clamp down on
reactivity.} The design of Atria was part anticipatory and part reactionary.
Both have risks. Anticipatory design may miss the mark. Reactionary
design may support too short-lived a target. By grounding ourselves in a long
history, we were able to judge any major addition in the context of long-term
concerns.

{\bf Jupyter notebooks impose additional design constraints, but open a wealth
of interaction opportunities.} Of our deployments, Jupyter has required the
most design changes and we foresee more as its use increases. While the
space constraints are greater, the cell-based interface could augment or
change how we develop and integrate interactive visualizations for use in this
data science space. Our current design depends on non-visualization cells and
scrolling to match functionality with our web version. This has worked well so
far, but further development of design guidelines for interactive notebook
environments are needed.

{\bf Rapid changes combined with multiple deployment targets incur a
maintenance burden.} While multiple deployments gave us many potential users
and their diverse viewpoints, they imposed a development burden on the
Visualization team. Each deployment is in a separate \texttt{git} branch and
requires some manual effort when applying changes. We plan to delve further
into how we can organize our code and development practices to decrease this
burden.

{\bf Both the visualization and the design study process aided our
collaborators in accomplishing their goals and helped establish a culture of
data review.} It can be difficult to discern whether it was a particular
visualization that led to an insight, or the fact that anyone was looking at
the data at all, especially in studies where visualization was not already in
the domain experts' workflow.

The integration of the Visualization team and the design process made data
collection and review a central priority. The dialogue between the teams and the
rapid response to data exposed data collection bugs or mis-assumptions early. As
seen in our regression case study (Section~\ref{sec:eval:deploy:als}), the
intervention of the design process worked in tandem with a specific analysis
problem to reach a solution.

We recommend further examination of the benefits of visualization as an
intervention, particularly with respect to developing best practices surrounding
a culture of data review. Based on our experience, we attribute our success to
the project organization from both sides. The rest of the project viewed the
Visualization aspects as first class deliverables. In turn, as key members of
the project, the Visualization team was also fully invested in other aspects of
the project. Although this investment brings certain risks, its rewards include
deep insights and impacts that are otherwise unavailable.

\section{Conclusion}
\label{sec:conclusion}

We presented a design study in the presence of the potential pitfalls
regarding lack of data availability or task recurrence.  The visualization
outcome and the insights it supports have not been the only benefit to our
domain collaborators. The design process itself and the integration with
the visualization efforts have been beneficial, especially as an avenue for
refining data collection and analysis practices. One of the goals of
the collaboration is to research what data needs to be collected for
asynchronous tasking runtimes. The evolution of the data has been in response
to intuition gained in analysis. The process has also resulted in rapid
verification of collected data and insights into current problems that we
anticipate will form a strong foundation for the on-going, long-term design
study.

Although we accepted some pitfalls as part of the project, several factors aided
us in managing them. Project organization was a large factor---teams had
respect for each others' expertise, met regularly, valued each others' time
and deadlines, and viewed the contributions of all teams as project
deliverables. The high level of participation resulted in a large corpus of
design data collected both by the Visualization team and the Runtime team.
This documentation was revisited frequently, both formally, through revising
the task analysis, and informally, to guide design efforts, avoiding ephemeral
needs. Acknowledging that the data and tasks were in flux, our technology
probe, Atria, satisfied those needs while keeping the focus on learning what 
data and tasks supported analysis rather than finalizing a tool design.

\acknowledgments{
We thank the members of the Phylanx Team, APEX Group, and Ste$||$ar groups.
This work was supported by the United States Department of Defense through DTIC Contract FA8075-14-D-0002-0007 and by the National Science Foundation under NSF III-1656958.
}

\bibliographystyle{abbrv-doi}

\bibliography{template}
\end{document}